\documentclass[twocolumn,showpacs,aip]{revtex4}
\usepackage{amssymb}
\usepackage{amsmath}
\usepackage{graphicx}
\usepackage{bm}
\usepackage{epstopdf}
\begin{document}

\title{Direct laser acceleration of electrons by tightly focused laser pulses.}

\author{Tianhong Wang$^1$, Vladimir Khudik$^1$$^,$$^2$, Alexey Arefiev$^3$ and Gennady Shvets$^1$}
 \affiliation{
   $^1$School of Applied and Engineering Physics, Cornell University, Ithaca, New York 14850, USA.
 \\$^2$Department of Physics and Institute for Fusion Studies, The University of Texas at Austin, Austin, Texas 78712, USA.
 \\$^3$Department of Mechanical Engineering and Center for Energy Research, University of California, San Diego, California 92093, USA.}

\date{\today}
\begin{abstract}
We present an analytical theory that reveals the importance of the longitudinal laser
electric field in the resonant acceleration of relativistic electrons by the tightly confined laser beam. It is shown that this field always counterworks to the laser transverse component and effectively decreases the final energy gain of electrons through direct laser acceleration mechanism. This effect
is demonstrated by carrying out the particle-in-cell simulations in the setup where the wakefield in
the plasma bubble is compensated by the longitudinal laser electric field experienced by the accelerated
electrons. The derived scalings and estimates are in good agreement with numerical simulations.


\end{abstract}

\maketitle
\section{Introduction}
As high intensity ($>10^{18}W/cm^{-2}$) laser beams propagate in an underdense plasma,  their radiation pressure expels electrons out of the propagation path.  Long beam  (with pulse duration $\tau$ much longer than $\omega_p^{-1}$, where $\omega_p =\sqrt{4\pi e^2n/m_e}$ is the plasma frequency, $n$ is the plasma density,  $m_e$ is the electron mass, and $-e$ is the electron charge) can support slowly evolving quasistationary ion channels in which the longitudinal  electric field is weak~\cite{pukhov_channel,mangles2005}. The laser wave accelerates electrons oscillating across the channel to the energy much higher than the energy gained by free electrons in the vacuum~\cite{Stupakov2001}. This acceleration regime is interesting for potential applications  of copious relativistic electrons as a compact X-ray sources~\cite{Park2006,Kneip2008, Stark}, and novel sources of energetic ions~\cite{Schollmeier2015}, neutrons~\cite{Pomerantz2014}, and positrons~\cite{Chen2015}.

Short beam ($\tau<\omega_p^{-1}$) generates an extremely nonlinear plasma wave blow-out structure known as a "plasma bubble"~\cite{pukhov_bubble} in which strong longitudinal electric field is capable of accelerating electrons injected at the bottom of the bubble up to multi-GeVs scale over short distance~\cite{leemans_1gev,leemans_4gev,wang_2gev,kim_3gev}.
The laser pulse  overlapping with relativistic electrons   additionally boosts their  energy  through direct laser acceleration (DLA)  mechanism~\cite{Gahn_DLA}. Recently it has been shown that the relativistic electrons injected in the front
portion of the plasma bubble in the decelerating phase of the plasma wave can also gain ultra-relativistic energy from the bubble-forming laser pulse through inverse ion-channel laser mechanism\cite{Kh_2018}.

During direct laser acceleration, the laser electric field pumps energy into transverse (betatron) oscillation which is redirected by the magnetic field into the longitudinal momentum.
The resonant interaction of electrons
with high-intensity laser wave is complicated by the fact that the Doppler shifted frequency of the wave $\omega_D = \omega_L(1-v_x/v_{ph})$, where
$ω_L$ is the wave frequency, $v_{ph}$ is its phase velocity and $v_x$ is the longitudinal particle velocity, only in average equals to the betatron frequency $\omega_{\beta}=\omega_p/\sqrt{2\gamma}$ of the electron oscillations  in the channel~\cite{shaw_ppcf}.  Strong non-linearity of the laser-particle interaction leads to an irregular (stochastic) motion of electrons in the ion channel~\cite{Krash_2018} and high sensitivity of the gained energy to the initial conditions.   

Since transverse laser wave fields are mainly responsible for DLA mechanism, DLA mechanism is analytically studied by approximating the laser pulse by a planar electromagnetic wave~\cite{Arefiev2012,Arefiev2014,Robinson,Arefiev2016,Khudik_2016,huang_2017}. 
 In this paper, we examine through an analytical theory and supporting PIC simulations the effect of a small but finite longitudinal electric field of the laser pulse in the DLA regime.  It is shown that this field can produce a significant deceleration effect that partially offsets the resonant acceleration by the transverse component of tightly focused laser pulses.

The remainder of the paper is organized as follows. In Sec.~II, we develop an analytical model by replicating the electromagnetic fields in the wave propagating in the ion channel.  We find in Sec.~III the scaling law for the work performed by the longitudinal component of the laser pulse. In Sec. IV we carry out particle-in-cell simulations of the resonant interaction of relativistic electrons with the  DLA laser pulse in the plasma bubble generated by the leading (pump) pulse~\cite{Xi_prl, zhang_ppcf,zhang_ppcf_2}.  Summary of the work is presented in Sec.~V.

\section{
Model}
\begin{figure}[t]
\centering
\includegraphics[width=.95\columnwidth]{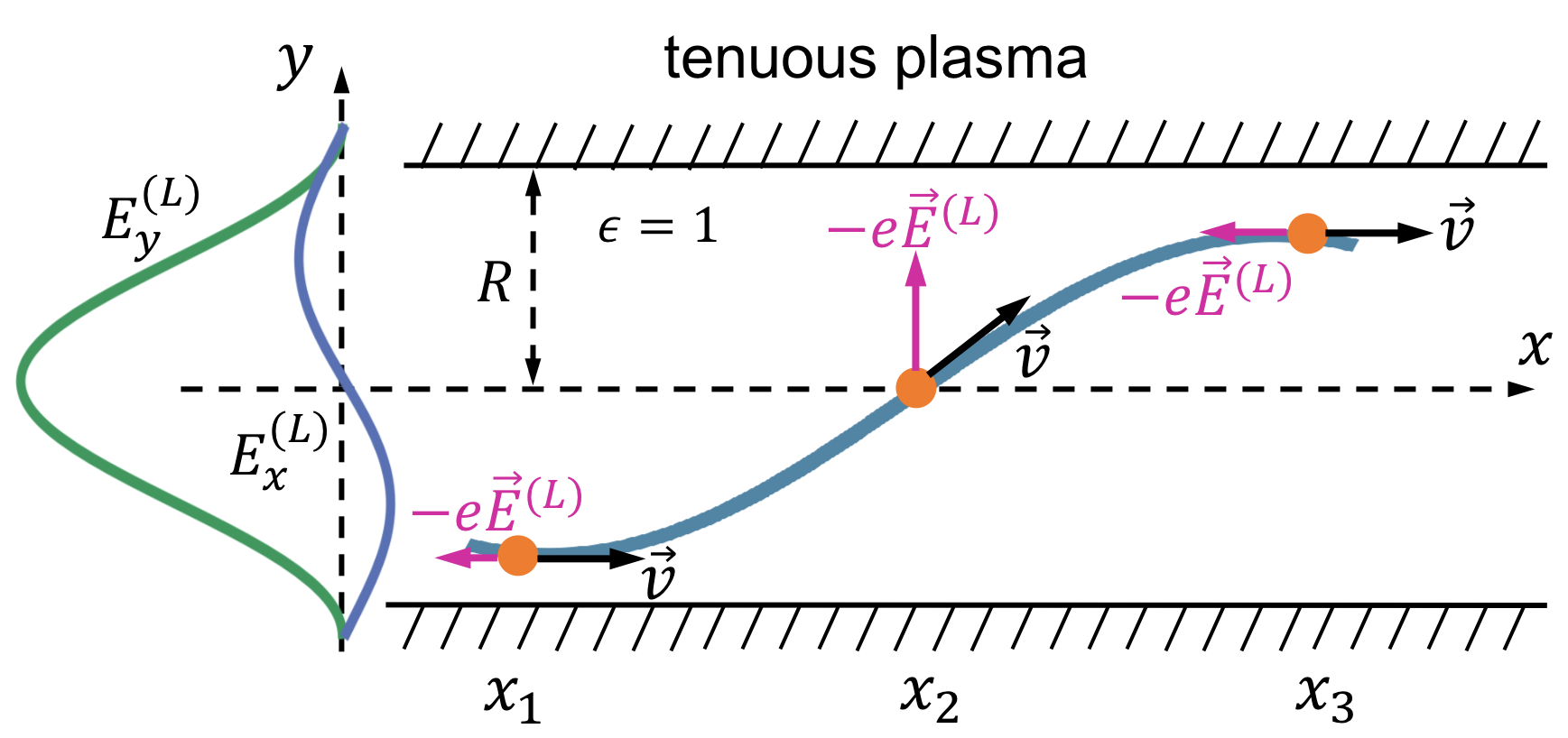}
\caption{Sketch of the electron motion 
and  electric fields of the laser pulse propagating in the ion channel. The electron quickly gains energy when it resonantly interacts 
with the laser wave. The phases of transverse and longitudinal components of the electric field ($E_y^{(L)}$   and  $E_x^{(L)}$)    are shifted by $\pi/2$
: $E_y^{(L)}=0$  at the electron turning points $x_1$ and $x_3$ while   $E_x^{(L)}$  has the same sign at these points.}
\label{fig:sketch}
\end{figure}

In order to capture the effect of laser longitudinal electric field, we consider a simplified setup which well approximates physics of the particle-wave interaction in the ion channels created by multi-picosecond laser beams, see  Fig.~\ref{fig:sketch}. 

   The ion channel is modeled under the following assumptions: it is created in tenuous plasma by full evacuation of electrons from a cylinder of the finite radius $R\gg \lambda_L$, where  $\lambda_L$ is the laser wavelength. We neglect the slow evolution of the channel considering it as a  quasistationary formation~\cite{Neda1}.  

The laser wave  propagates    in the $x$-direction   aligned with the channel axis.  It is linearly  polarized in $y$-direction so that   $E_y^{(L)}$  is the main component of the laser electric field.
%
In the cylindrical channel, there are only two components of the electric field  $E_y^{(L)}$, $E_x^{(L)}$ and one component of the magnetic field $B_z^{(L)}$) do not vanish in the main plane $x-y$ depicted in Fig.~1. 
The  radial electrostatic field in the channel is approximated as $E_y =  m\omega_p^2 y/2e$.
The dynamics of  relativistic electrons     is then described in this model by:
\begin{eqnarray}
&&\frac{d p_x}{dt} =-eE_x^{(L)}-\frac{e}{c}v_y B_z^{(L)}, \label{eq:EqM_N1}\\
&&\frac{d p_{y}}{dt} = -\frac{1}{2}m\omega_p^2 y-e\Big(E_y^{(L)}-\frac{v_x}{c}B_z^{(L)}\Big), \label{eq:EqM_N2}\\
&&\frac{d x}{dt}={{v_x}}=\frac{{p_x}}{m\gamma},\quad \frac{d y}{dt}={{v_y}}=\frac{{p_y}}{m\gamma}
\label{eq:EqM_N3}
\end{eqnarray}
where $\gamma=(1+{\bf{p}}^2/m^2c^2)^{1/2}$ is the electron relativistic factor, and   $c$ is the speed of light. Compared with previous models, where the laser wave is considered to be planar,  these equations contain the longitudinal component of the laser electric field which can profoundly impact the energy gain of resonantly accelerated electrons. 

We assume that  $E_y^{(L)}\propto \cos{\phi}$, where $\phi= \omega_L(x/v_{\rm ph} - t)\equiv k_xx -\omega_L t$ is the phase of the laser wave, 
and $k_x= \omega_L/v_{\rm ph}$ is the $x$-component of the laser wavenumber. Since heavy ions in the channel are considered to be immobile, this field satisfies the wave equation in the vacuum:
\begin{eqnarray}
\nabla_{\perp}^2 E_y^{(L)}=-k_{\perp}^2E_y^{(L)}.
\label{eq:Eq_EyL}
\end{eqnarray}
where $k_{\perp}=\sqrt{k_L^2-k_x^2}$ and $k_L=\omega_L/c$.

The $z$-component of the electric field does not vanish in the $y-z$ plane and can be estimated in the tenuous plasma as $E_z^{(L)}\lesssim (\omega_p/\omega_L)^2E_y^{(L)}\ll E_y^{(L)}$.
Then, the longitudinal component $E_x^{(L)}$ is found from  Poisson equation 
\begin{eqnarray}
\frac{\partial E_x^{(L)}}{\partial x}+\frac{\partial E_y^{(L)}}{\partial y}=0,
\label{eq:Eq_ExL}
\end{eqnarray}
 in which a small term $\partial E_z^{(L)}/\partial z\ll \partial E_y^{(L)}/\partial y$ is omitted, and the $z$-component of the magnetic field is found from the Ampere-Maxwell equation 
\begin{eqnarray}
-\frac{1}{c}\frac{\partial B_z^{(L)}}{\partial t}=\frac{\partial E_y^{(L)}}{\partial x}-\frac{\partial E_x^{(L)}}{\partial y}.
\label{eq:Eq_BzL}
\end{eqnarray}
The solution of  Eqs.~(\ref{eq:Eq_EyL}) - (\ref{eq:Eq_BzL}) for the linearly polarized laser wave is given by
\begin{eqnarray}
&&E_y^{(L)} = E_0 J_0(k_{\perp}y)\cos{\phi},
\label{eq:Eq_Eyy}\\
&&E_x^{(L)} =\frac{k_{\perp}}{k_x}E_0J_1(k_{\perp}y)\sin{\phi} , \label{eq:Eq_Exx}\\
&&B_z^{(L)} = \frac{c}{v_{\rm ph}}E_0 \Big[ J_0(k_{\perp}y)+\frac{k_{\perp}^2}{k_x^2} J'_1(k_{\perp}y)\Big]\cos{\phi},
\label{eq:Eq_Bzz}
\end{eqnarray}
where $E_0=a_0(m c\omega_L/e)$ is the amplitude of the laser electric  field, $a_0$ is the dimensionless amplitude of the vector potential, $J_0(\xi)$ and $J_1(\xi)$ are Bessel functions of the first kind and $J_1'(\xi)\equiv dJ_1(\xi)/d\xi$. 

The transverse wavenumber $k_{\perp}$ for fundamental mode of the linearly polarized laser wave~\cite{okamoto} can be estimated by   setting outside the channel  the main component of the electric field (\ref{eq:Eq_Eyy}) to zero:
\begin{eqnarray}
J_0(k_{\perp}R_*)=0, \quad k_{\perp}\approx 2.4/R_*,
\label{eq:Kperp}
\end{eqnarray}
where $R_*\equiv R+\delta_s$ and $\delta_s=c/\omega_p$ is accounted for the finite skin depth. 
Then we estimate $k_x=(k_L^2-k_{\perp}^2)^{1/2}\approx k_L[1-2.88/(k_L^2R_*^2)]$ and  
\begin{eqnarray}
v_{\rm ph}\approx c[1+2.88/(k_L^2R^2)].
\label{eq:v_ph}
\end{eqnarray}

\begin{figure}[t]
\centering
\includegraphics[height=0.14\textheight, width=.95\columnwidth]{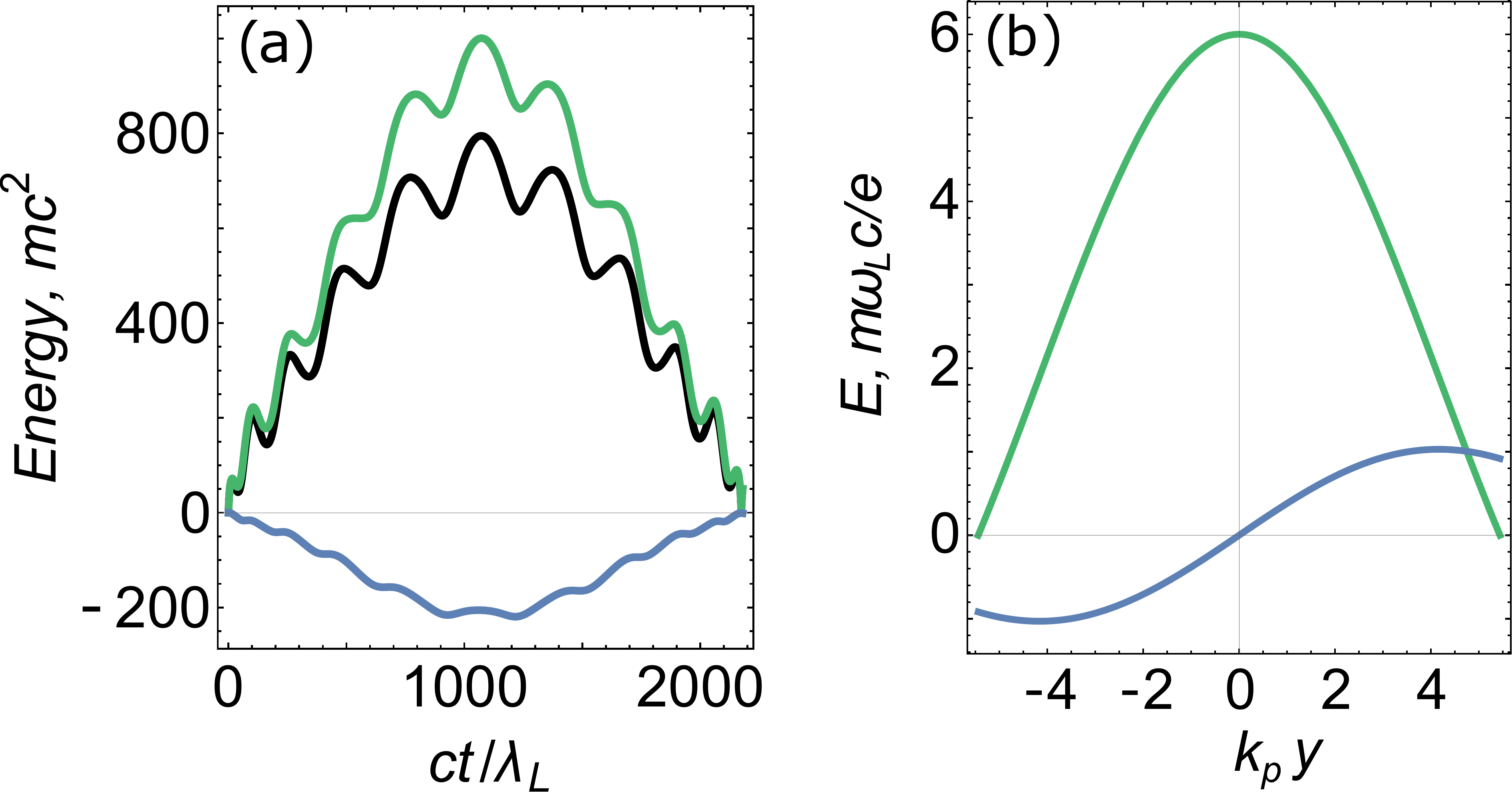}
\caption{ The energy gain of electron resonantly interacting with the laser wave  confined in the ion channel. (a) The time dependence of the  electron energy  (black) and the  works  performed by the transverse  and longitudinal  laser fields (green and blue, respectively).  The electron is initially placed at rest on the channel axis at $\phi|_{t=0}=\frac{\pi}{2}$. The laser-plasma parameters are: $a_0=6$, $\omega_p/\omega_L=1/15$, $k_pR_*\approx 5.44$. (b) Transverse and longitudinal laser fields in the ion channel:  $E_y^{(L)}(y)$ (brown) and $10\times E_x^{(L)}(y)$ (blue); the phase of these fields is shifted by $\pi/2$. }\label{fig:channel_accel}
\end{figure}
 Equations~(\ref{eq:EqM_N1}) - (\ref{eq:EqM_N3}) complemented by the expressions for fields (\ref{eq:Eq_Eyy}) - (\ref{eq:Eq_Bzz}) fully determine relativistic  dynamics of   electrons under action of the laser  wave  confined in the ion channel. 

Typical energy gain during particle-wave resonant interaction is shown in Fig.~\ref{fig:channel_accel}. 
First noteworthy feature of such an interaction is that  the different components of the laser electric field work in opposite way: while    $E_y^{(L)}$ pumps energy into electron oscillations  effectively accelerating  
 particle,   $E_x^{(L)}$ returns part of this energy to the laser wave   decreasing the resulting energy gain,  see Fig.~\ref{fig:channel_accel}(a). The other striking feature is that although the longitudinal   component is much less than the  transverse one [$E_x^{(L)}/E_y^{(L)}\sim 1/60$ in Fig.~\ref{fig:channel_accel}(b)], the corresponding longitudinal and transverse works  are not so different: $|W_{||}|/W_{\perp}\sim 1/5$ (and  $|W_{||}|/mc^2\gamma\sim 1/4$). 

\section{Resonant DLA in  ion channels}
In this section, we  examine the motion of relativistic electrons governed by Eqs.~(\ref{eq:EqM_N1}) - (\ref{eq:EqM_N3})  
and
derive an analytical expression for the work done by  longitudinal  component of the laser beam.
    
\subsection{Phase coupling}
We start our analysis by noting that Hamiltonian $H$ of particle motion  depends on $x$ and $t$ only through combination $x-v_{\rm ph}t$, and hence $H-v_{\rm ph}P_x$, where $P_x$ is the $x$-component of the canonical momentum, is conserved.
 Therefore,  equations~(1) - (3) have the following integral of motion:
\begin{eqnarray}
\gamma-\frac{v_{\rm ph}}{c}\Big(\frac{p_x}{mc}-a_x\Big) +\frac{1}{4}k_p^2y^2=I_0=const, 
\label{eq:integral0}
\end{eqnarray}
where $a_x=-(k_{\perp}/{k_x})a_0J_1(k_{\perp}y)\cos\phi$ is the $x$-component of the dimensionless vector potential of the laser wave, and the constant $I_0$ is  determined by the initial position and momentum of the particle.
Equation~(\ref{eq:integral0}) can be  simplified for the luminal wave ($v_{\rm ph}=c$) and electrons moving with relativistic speed along channel axis ($p_x\gg p_y\gg 1$) 
\begin{eqnarray}
&&\frac{1}{2c^2}\gamma v_y^2 +\frac{1}{4}k_p^2y^2=I_0-a_x, 
\label{eq:integral}
\end{eqnarray}
Assuming that $I_0\gg a_x$, we find that $y\approx 2(I_0)^{1/2}k_p^{-1}\sin\psi$, $v_y\approx c(2I_0/\gamma)^{1/2}\cos\psi$, and $d \psi/dt\equiv \omega_{\beta}=\omega_p/(2\gamma)^{1/2}$, where $\psi$ is the phase of betatron oscillations. In this regime the time-derivative of the wave phase   can be expressed in terms of the phase of betatron oscillations: $d\phi/dt=-\omega_D$, where $\omega_D =\omega_L(1-v_x/c)\approx \omega_L(1-v_x^2/c^2)/2\approx \omega_Lv_y^2/2c^2=\omega_L(I_0/\gamma)\cos^2\psi$. Assuming that change of the relativistic factor is small during one betatron period, we can find that $d\phi/d\psi= -2(\langle\omega_D\rangle /\omega_{\beta})\cos^2\psi$ and hence the wave phase and the phase of betatron oscillations of the relativistic electron are  coupled by the relation 
\begin{eqnarray}
&&\phi =\theta-(\langle\omega_D\rangle /\omega_{\beta})\Big(\psi+\frac{1}{2}\sin2\psi\Big), 
\label{eq:psi}
\end{eqnarray}   
where $\theta$ is the phase mismatch which approximately constant during one betatron oscillation,   and $\langle\omega_D\rangle=\omega_L/(2 I_0\gamma)$.
\subsection{Scaling law}
The relationship between the longitudinal and transverse works, $W_{||}$ and $W_{\perp}$, and the relativistic factor $\gamma$ during resonant interaction of relativistic electrons co-propagating with  laser wave in the ion channel
can be easily found for small amplitude of betatron oscillations.  In this case
the  components of the laser electric field can be approximated as $E_x\approx  \frac{1}{2}(k_{\perp}/k_x)  k_{\perp}y E_0\sin\phi$ and  $E_y^{(L)}\approx E_0\cos\phi$. Then 
  expressions for corresponding works (normalized to $mc^2$) take the following form:  
\begin{eqnarray}
\frac{d {W}_{||}}{dt}=-\frac{eE_x^{(L)}v_x}{mc^2}\approx -eE_0c\frac{k_{\perp}^2}{k_Lk_p}  I_0^{1/2}\sin\phi \sin \psi, 
\label{eq:Wx}
\\
\,\,\frac{d {W}_{\perp}}{dt}=-\frac{eE_y^{(L)}v_y}{mc^2}\approx -eE_0c\sqrt{\frac{2}{\gamma}}I_0^{1/2}\cos\phi \cos \psi,\label{eq:Wy}
\end{eqnarray}

Assuming that $\gamma$ is approximately constant during one betatron period, we can use Eq.~(\ref{eq:psi}) to  average the trigonometric functions in r.h.s. of  Eqs.~(\ref{eq:Wx}) and (\ref{eq:Wy})  over betatron oscillations. The result is especially simple   near the resonance $\omega_{\beta}\approx\langle\omega_D\rangle$~\cite{Davoine2014}: $\langle \sin\phi \sin \psi\rangle=\alpha_{||}\cos \theta$ and $\langle \cos\phi \cos \psi\rangle=\alpha_{\perp}\cos \theta$, where $\alpha_{||}=-0.59$ and $\alpha_{\perp}=0.348$. Since $\alpha_{||}$ is negative,  the longitudinal component of the laser electric field always  works in opposite way compared to the transverse component. Note that during the resonant motion depicted in Fig.~1,  the phase mismatch is $\theta=-\pi$, and phases of the betatron oscillations and laser wave at the position $x=x_1$ are $\psi|_{x=x_1}=\phi|_{x=x_1}=-\frac{\pi}{2}$.

\begin{figure}[t]
\centering
\includegraphics[width=.95\columnwidth]{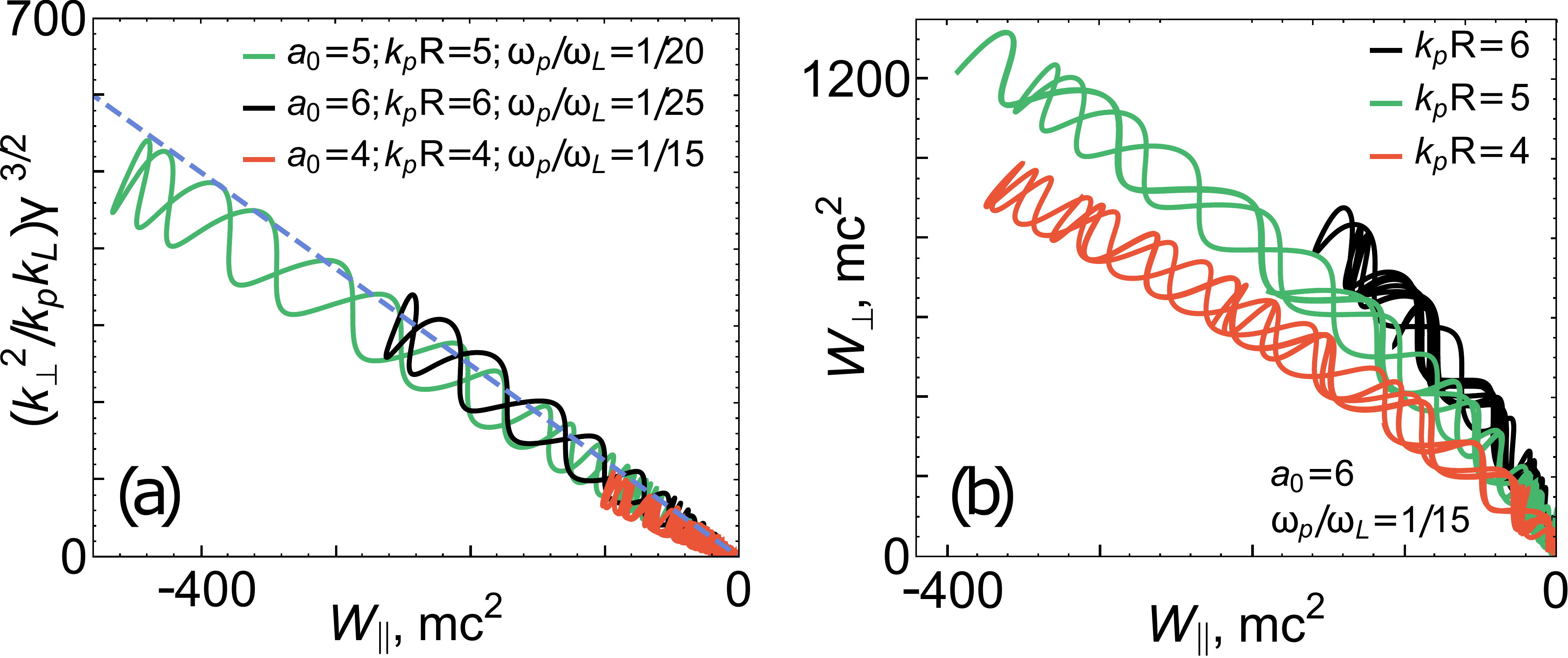}
\caption{  (a) Relationship between  the  relativistic factor and the longitudinal work performed by the laser wave  at different  laser-plasma parameters. Dashed line corresponds to the scaling law~(\ref{eq:Wx_Wy}). (b) Dependence $W_{\perp}$ on ${W}_{||}$ at different $R$ and the same laser intensity and plasma frequency: the ratio $|{W}_{||}/W_{\perp}|$ increases with decrease of the channel radius. In all cases,  electrons are initially placed at rest on the channel axis at $\phi|_{t=0}=\frac{\pi}{2}$. }\label{fig:scaling}
\end{figure}

For sake of simplicity, we  neglect the change of the numerical factors $\alpha_{||}$ and $\alpha_{\perp}$ caused by the detuning from the resonance.
 Then we  reduce averaged  Eqs.~(\ref{eq:Wx}) and (\ref{eq:Wy}) to the one equation  
\begin{eqnarray}
{d {W}_{||}}/{d {W}_{\perp}}=-1.2\sqrt{\gamma}{k_{\perp}^2}/({k_Lk_p}),
\label{eq:Wx_over_Wy}
\end{eqnarray}
from which we find  the upper limit for  electron relativistic factor   $\gamma_{up}\equiv  [k_Lk_p/1.2k_{\perp}^2]^2$ and the scaling law for the $W_{||}$ (in the lowest order of $1/\gamma_{up}$): 
\begin{eqnarray}
\frac{{W}_{||}}{mc^2}\approx -\frac{2}{3\sqrt{\gamma_{up}}} \gamma^{3/2}, 
\label{eq:Wx_Wy}
\end{eqnarray}
This formula  does not depend on the initial conditions (coordinates and momenta of electrons), the wave amplitude and the  final energy  gain. Despite many approximations,  the scaling law~(\ref{eq:Wx_Wy}) describes very well the relation between $W_{||}$ and $\gamma$ obtained from exact integration of Eqs.~(\ref{eq:EqM_N1}) - (\ref{eq:EqM_N3}) with  fields~(\ref{eq:Eq_Eyy}) - (\ref{eq:Eq_Bzz}), see Fig.~\ref{fig:scaling}(a). Note that $\gamma_{up}\propto k_L^2k_p^2R^4$ and therefore a negative role of the longitudinal electric field quickly increases with decrease of the channel radius, see Fig.~\ref{fig:scaling}~(b). 

Expressions~(\ref{eq:Wy}) - (\ref{eq:Wx}) for transverse and longitudinal work show
 that the large ratio $E_y^{(L)}/E_x^{(L)}$  is partially counterbalanced for  relativistically moving particles  by the small ratio $v_y/v_x\sim 1/\sqrt{\gamma}$. This  results in moderate difference between $W_{\perp}$ and $W_{||}$ (or  $\gamma$ and $W_{||}$). Effective acceleration (or deceleration)  by the transverse force is accompanied by deceleration (or acceleration) by the longitudinal force.  In Appendix, we show that the energy gained by electrons resonantly interacting with the laser pulse confined in the ion channel is always smaller than the energy gained from the planar wave provided that amplitudes and phase velocities are the same. 

Since the scaling law~(\ref{eq:Wx_Wy}) is derived by using a very crude approximation for the $y$-dependence of laser fields, it is not sensitive to specifics of field profiles across the channel. In particular, it can be used for planar channels  after replacing 
$\gamma_{up}\rightarrow [k_Lk_p/1.7k_{\perp}^2]^2$.

\section{\label{sec1} PIC simulations}

The developed model assumes that accelerating field in the long ion channel is negligibly small. 
In this section, we design a numerical experiment that clearly the longitudinal laser electric field can significantly change the energy balance of DLA electrons accelerated simultaneously by the laser pulse and the wakefield in the small plasma bubble. The electron dynamics is modeled in realistic 3D geometry using the first-principle  self-consistent relativistic 3D PIC code VLPL~\cite{Pukhov_code}
and quasistatic in-home PIC  code (QS-DLA)~\cite{TH_2017}.   The latter can be used as a convenient tool for analyzing the contribution to the electron energy from wake- and laser fields.

\begin{figure}[b]
\centering
  \includegraphics[width=0.95\columnwidth]{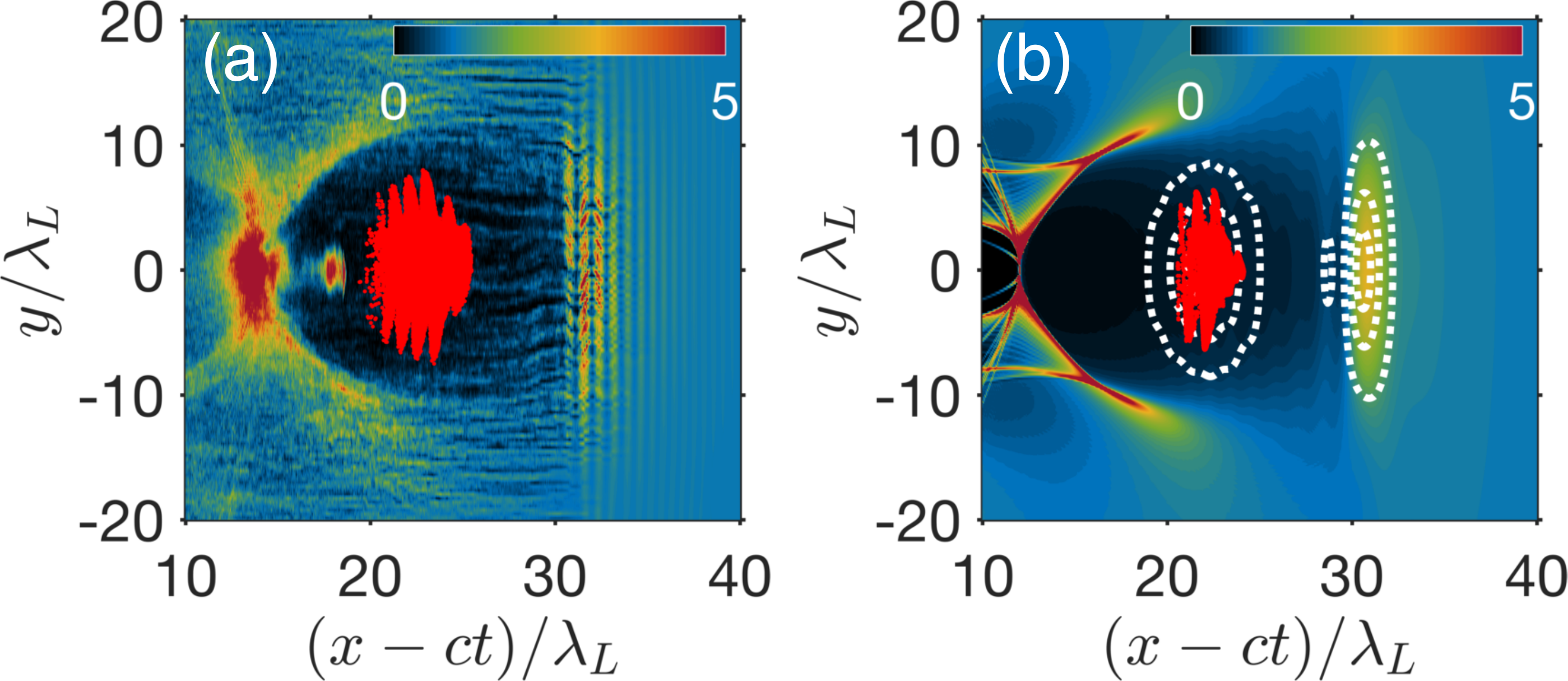}\\
\caption{(a) VLPL and (b) QS-DLA: Plasma bubble  at  propagation distance  $x=1.15 mm$; externally injected electrons are marked by red points, and pump and DLA pulse contours are plotted with white dashed lines. }
\label{Fig4}
\end{figure}  

We launch  two (pump and DLA) laser pulses separated by distance 10$\mu$m in the tenuous plasma  with density $n_0=7.7\times10^{18}\rm cm^{-3}$.  The leading  pump pulse with the wavelength $\lambda_L=0.8\mu m$, power $P=21TW$, duration $\tau_{\rm pump}=16.6 fs$,  spot size $w_{\rm pump}=8.7\mu m$,  and dimensionless vector potential $a_0=2.9$  blows out electrons  on its path creating  a plasma bubble.
 The following DLA pulse with the same wavelength and power, duration $\tau_{\rm DLA}=9.4fs$, spot size $w_{\rm DLA}=5.5\mu m.$, and dimensionless vector potential $a_0=4.6$ is placed initially near the back of the bubble. 

A short electron bunch with duration $\tau_b=4fs$, transverse size $3\mu m$ and small density $n_b=6.7\times10^{15}\rm cm^{-3}$   is externally injected with initial momentum $p_x=15mc$ in the $x-y$ plane  near the center of  the DLA pulse.

\begin{figure}[b]
\centering
  \includegraphics[width=0.95\columnwidth]{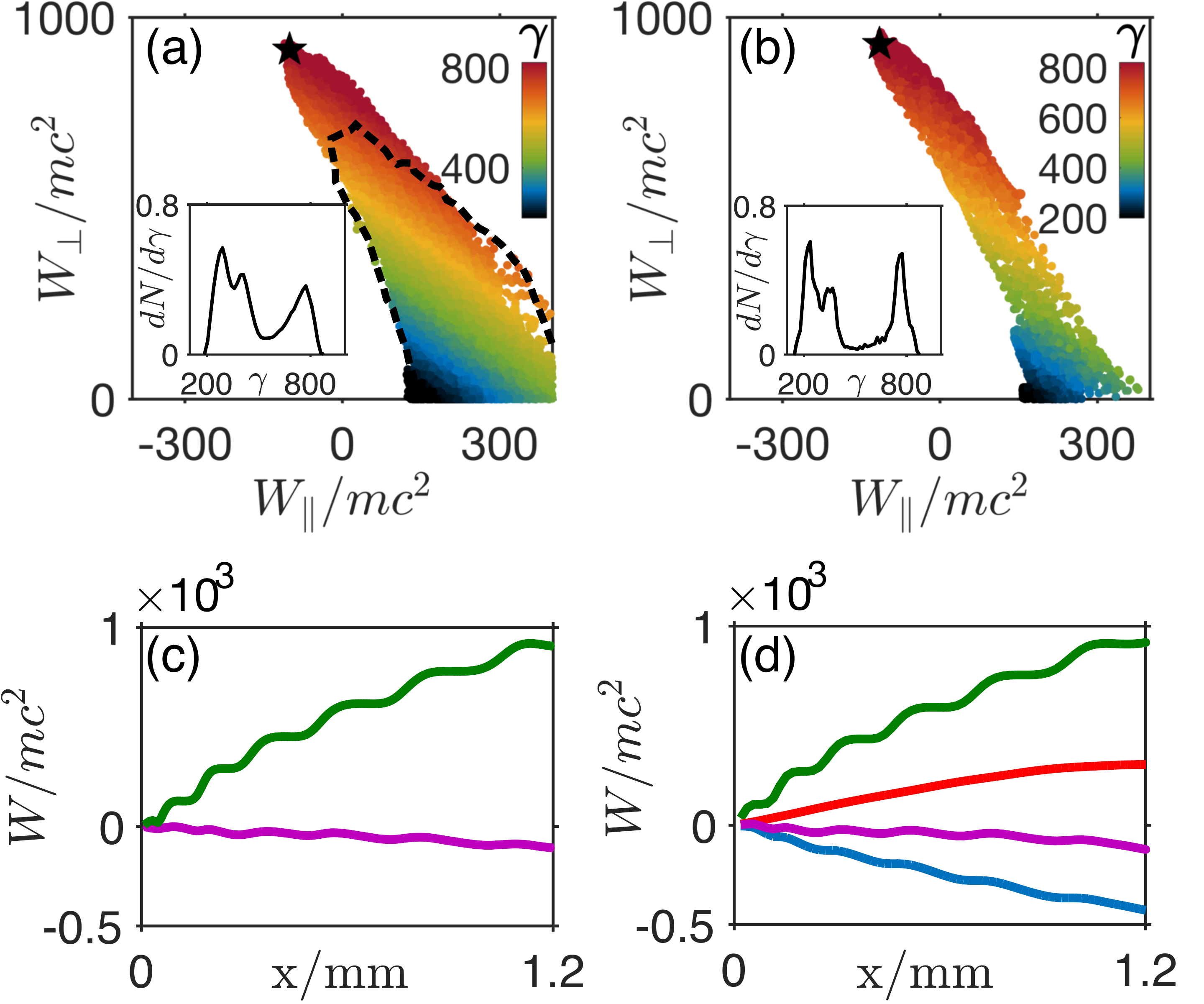}\\
\caption{(a) VLPL and (b) QS-DLA: Phase space ($W_{||}, W_{\perp}$) of injected electrons color coded by their relativistic factor $\gamma$  at the propagation distance  $x=1.15 mm$. Insets  show energy spectrum of  these electrons. Dashed black curve in (a) encloses a phase space   of particles obtained from VLPL simulations at twice lower resolution ($c\Delta t=\lambda/50$).  (c) VLPL and (d) QS-DLA: Transverse  $W_{\perp}$ (green)  and  longitudinal $W_{||}$ (magenta) work   as a function of the propagation distance $x$ for the  energetic electrons selected by 'black star' in panels (a) and (b).   The total longitudinal work $W_{||}$   is decomposed in (d) into the works $W_{||}^{(wake)}$ (red) and $W_{||}^{(L)}$ (blue) done by the wake  and   the laser, respectively. }
\label{Fig4_2}
\end{figure} 

 The injected electrons co-propagate with the DLA laser pulse simultaneously experiencing acceleration by the laser wave and the bubble wakefield. They approximately reach the bubble
center at the dephasing distance $x=1.15 mm$, see   Fig.~\ref{Fig4} depicting these electrons and the plasma electron density. 

As seen from panels (a) and (b) of this figure, the results of the VLPL and QS-DLA simulations are very similar to each other except for some insignificant differences. 
Specifically,  VLPL simulations reproduce correctly self-injection mechanism~\cite{kalmykov}.  Self-injected electrons are trapped in the plasma bubble during its evolution and localized near the bubble axis behind the cloud of externally injected electrons. The simulations also exhibit some difference in spatial dimensions of the externally injected bunch when it approaches the plasma bubble center. However, in both cases the high energy electrons are positioned at the back of the bunch while the low energy electrons are in bunch front~\cite{zhang_ppcf}. 

\begin{figure}[t!!!]
\centering
\includegraphics[width=0.95\columnwidth]
    {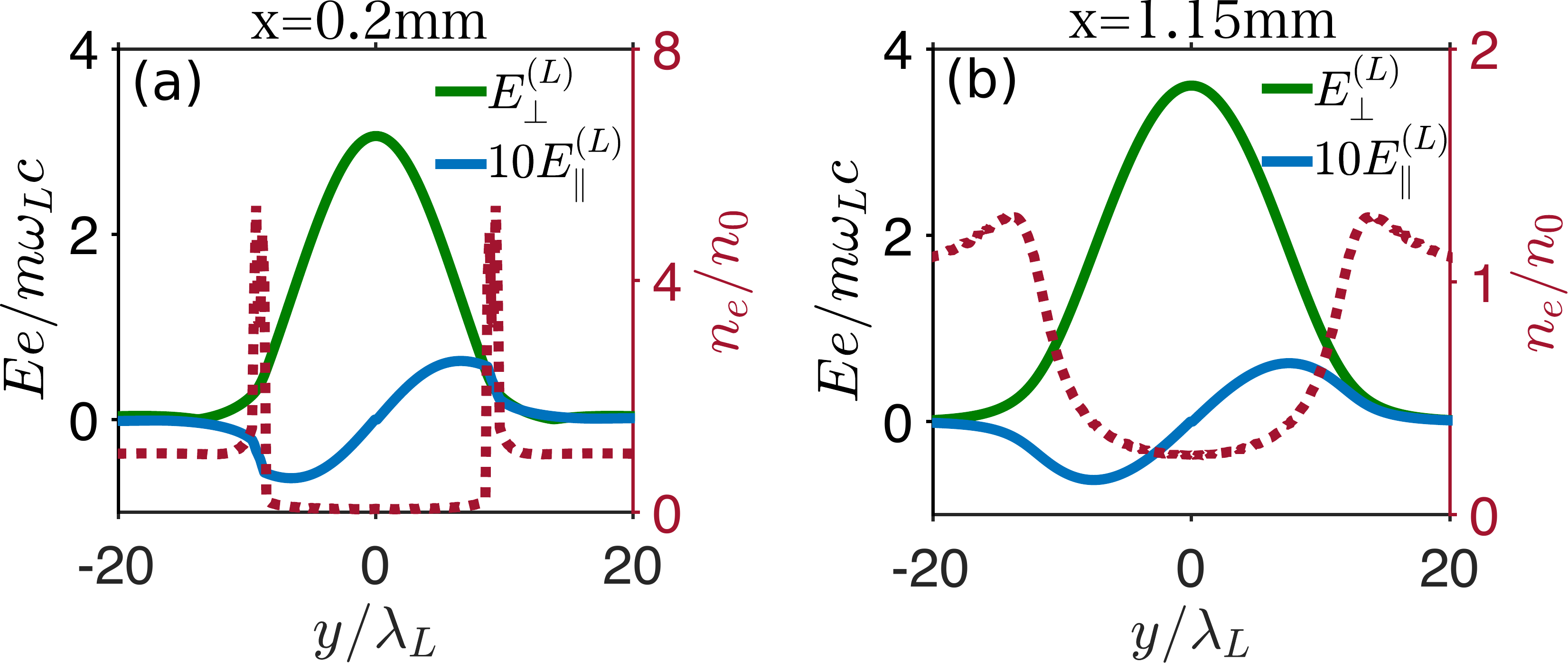}\\
\caption{  Dependence of $E_y^{(L)}$ (green), $E_x^{(L)}$  (blue) and  $\chi\equiv\langle n_e/\gamma \rangle$ (black) on  $y$ at the same propagation distances.   Ratio $|E_y^{(L)}/E_x^{(L)}|$ is approximately equal to $45$ at $x=0.2 mm$ and $60$ at $x=1.15 mm$.  
  }
\label{Fig5}
\end{figure}
More detailed characteristics of the injected bunch are presented in Fig.~\ref{Fig4_2}. The phase space ($W_{||}, W_{\perp}$) of particles from VLPL (a) and QS-DLA (b) simulations   indicates that the longitudinal work   for significant fraction of energetic electrons is negative: $W_{||}=-e\int E_xv_xdt<0$. If the longitudinal laser electric field were left out of the consideration, one could reach paradoxical conclusion that the wakefield  in the plasma bubble decelerates electrons despite the fact that they are positioned in the accelerating phase of the wakefield wave.

In reality,  a considerable fraction of the energy gained through the work of the transverse component returns to the laser wave through the negative work of its longitudinal component.  This is clarified in Fig.~\ref{Fig4_2}(c) and (d): In panel (c) we present the change of $W_{\perp}$ and $W_{\parallel}$ with propagation distance for selected particle in VLPL simulations, and in panel (d) we use advantage of quasistatic code and decompose $W_{\parallel}$ into    contributions of the wake and longitudinal laser electric fields. It turns out that in this regime the laser longitudinal component returns to the laser wave approximately half of the energy gained from the work of the transverse component.

In contrast to the model consideration, the conditions of resonant particle-wave interaction in the plasma bubble are constantly changing with time as DLA pulse advances to the bubble center: its amplitude decreases while spot size and the ratio between transverse and longitudinal electric fields increase, see Fig.~\ref{Fig5} (a) and (b).  
 Despite that, the ratio  between $W^{(L)}_{||}$ and $W_{\perp}^{(L)}$  in Fig.~\ref{Fig4_2}(c) is reasonably well described by scaling law~(\ref{eq:Wx_Wy})  if one uses averaged bubble radius 
$R\approx 10.5\lambda_L$.  

Concluding this section, note that in order to accurately reproduce a resonant nature of  DLA mechanism by the first principles PIC code, the time step (and the longitudinal cell size) ought to be small. When we decrease resolution of simulations (from $c\Delta t= \lambda/100$ to $c\Delta t= \lambda/50$), the electron  distribution in the phase space 
$W_{||}, W_{\perp}$ has changed significantly, see the area inside of the dashed curve in Fig.~\ref{Fig4_2}(a).

\section{Conclusion}
We have found that the longitudinal component of the laser electric field can make a profound impact on the energy balance of accelerated (or decelerated) electrons. During resonant direct laser acceleration of particles in ion channels or in plasma bubbles, the transverse and longitudinal components of the laser electric field effectively act in an opposite way: if one component accelerates a particle, the other decelerates it. Since a difference in the field components ($E_x^{(L)}\ll E_y^{(L)}$)  is partially counterbalanced by the difference in the velocity components ($v_x\gg v_y$), the corresponding works $W_{||}^{(L)}$ and $W_{\perp}^{(L)}$ can be comparable when DLA  pulses propagate in narrow channels or plasma bubbles.    The scaling law~(\ref{eq:Wx_Wy})  gives a good estimate for $W_{||}$ and for the upper limit of the electron energy gained through DLA mechanism.

 \section{Acknowledgments}
This work was supported by DOE grants DE-SC0007889 and
DE-SC0010622, and by AFOSR grant FA9550-14-1-0045. The authors thank the
Texas Advanced Computing Center (TACC) at The University of Texas at Austin for providing HPC resources.

\appendix
\section{Energy gain from the confined laser pulse propagating in the ion channel}

In this Appendix, we show that the energy gained by electrons resonantly interacting with the laser pulse confined in the ion channel is always smaller the energy gained from the planar wave provided that amplitudes and phase velocities are the same. 

\begin{figure}[b]
\centering
  \includegraphics[width=1.0\columnwidth]{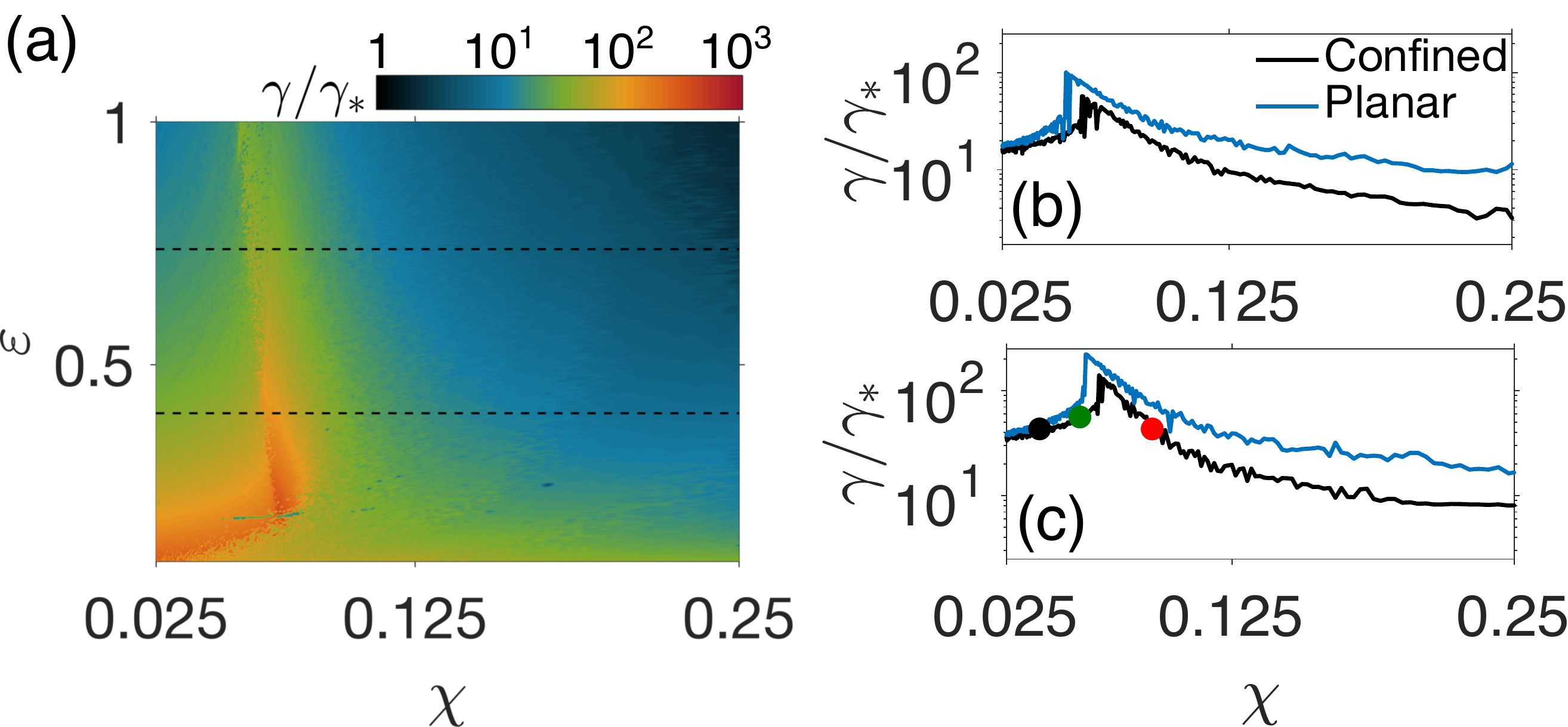}
\caption{ (a) Maximum energy as a function of the $\mathcal{E}\propto a_0$ and $\chi\propto R^{-2}$; $\gamma$ is normalized to $\gamma_*=a_0^2/2$, and dashed lines correspond to $\mathcal{E}=0.74$ and $\mathcal{E}=0.4$.  Electrons are initially placed at rest on the channel
axis at $\phi|_{t=0}=\pi/2$. (b) Comparison of the maximum energy gained by electrons from the confined and planar   waves along dashed lines in panel (a). Three colored circles (black, green, and red) correspond to three curves (black, green, and red) at the laser-plasma parameters used in Fig.~3.}
\label{Fig6}
\end{figure}

The confined laser pulse propagates in the ion channel with phase velocity $v_{ph}>c$. This makes analytical consideration quite complicated. However, when $v_{ph}-c\ll c$ ($k_LR_*>>1$), equations of motion  written in the dimensionless form depend only on two universal parameters  $\mathcal{E}$ and $\chi$, which combine laser-plasma properties and electron’s initial conditions:
\begin{eqnarray}
\mathcal{E}=a_0(\omega_p/\omega_L)(\tilde{v}_{ph}/c)I_0^{-3/2}, 
\label{strength}
\\
\chi=(I_0/\tilde{v}_{ph}^2)(\tilde{v}_{ph}-1)/2\omega_p^2/\omega_L^2),
,\label{dispersion_1}
\end{eqnarray}
 where $\tilde{v}_{ph}={v}_{ph}/c$. Using Eq.~(\ref{eq:v_ph}), we can express  the parameter $\chi$ characterizing  superluminosity  of the wave through the radius of the ion channel:
\begin{eqnarray}
\chi\approx\frac{1.44I_0}{k_p^2R^2}
\label{dispersion_2}
\end{eqnarray}
We integrate Eqs.~(\ref{eq:EqM_N1}) - (\ref{eq:EqM_N3})  
with fields (\ref{eq:Eq_Eyy}) - (\ref{eq:Eq_Bzz}) and find the  maximum energy  gained by electrons from the confined wave  as a function of the $\mathcal{E}$ and $\chi\propto R^{-2}$, see Fig.~\ref{Fig6}(a).
In fugure~\ref{Fig6}(b) and (c) we compare maximum energy gained by electrons from confined and planar wave. The latter is defined as~\cite{Kh_2018} 
\begin{eqnarray}
&&E_y^{(L)} = E_0\cos{\phi},
\label{eq:Eq_Eyy__}\\ 
&&E_x^{(L)} =0,
\label{eq:Eq_Exx__}\\
&&B_z^{(L)} = \frac{c}{v_{\rm ph}}E_0 \cos{\phi}.
\label{eq:Eq_Bzz__}
\end{eqnarray}
As expected, the difference between acceleration by the confined and planar wave becomes large as the channel radius decreases (that is, as $\chi$ increases).
\newpage


\end{document}